\documentclass[twocolumn]{aastex701}

\newcommand{\sdss}{\ref{fig:SDSSJ}}
\newcommand{\masx}{\ref{fig:2MASXJ}}

\newcommand{\p}{$\pm$}

\begin{document}

\title{Ongoing Morphological Transformation from Elliptical to Spiral Galaxies Induced by Gas Accretion from Neighboring Galaxies}

\author[orcid=0009-0001-3390-5151, sname='Yun']{Gene Yun}
\affiliation{Department of Physics and Astronomy, Seoul National University, 1 Gwanak-ro, Gwanak-gu, Seoul 08826, Republic of Korea} 
\email[show]{geneyun64@gmail.com}

\author[orcid=0009-0003-9748-4194, sname='Im']{Sang Hyeok Im}
\affiliation{Department of Physics and Astronomy, Seoul National University, 1 Gwanak-ro, Gwanak-gu, Seoul 08826, Republic of Korea}
\affiliation{Korea Institute for Advanced Study, 85 Hoegi-ro, Dongdaemun-gu, Seoul 02455, Republic of Korea}
\email{sanghyeok.im97@gmail.com}

\author[orcid=0000-0003-3428-7612, sname='Hwang']{Ho Seong Hwang}
\affiliation{Department of Physics and Astronomy, Seoul National University, 1 Gwanak-ro, Gwanak-gu, Seoul 08826, Republic of Korea}
\affiliation{SNU Astronomy Research Center, Seoul National University, 1 Gwanak-ro, Gwanak-gu, Seoul 08826, Republic of Korea}
\affiliation{Institute for Data Innovation in Science, Seoul National University, Seoul 08826, Republic of Korea}
\email[show]{hhwang@astro.snu.ac.kr}

\author[orcid=0000-0001-9521-6397, sname='Park']{Changbom Park}
\affiliation{Korea Institute for Advanced Study, 85 Hoegi-ro, Dongdaemun-gu, Seoul 02455, Republic of Korea}
\email{cbp@kias.re.kr}

\author[orcid=0009-0002-9878-1126, sname='Bahk']{Hyeonguk Bahk}
\affiliation{Department of Physics and Astronomy, Seoul National University, 1 Gwanak-ro, Gwanak-gu, Seoul 08826, Republic of Korea}
\email{spica095@snu.ac.kr}

%\correspondingauthor{Ho Seong Hwang}
%\email{hhwang@astro.snu.ac.kr}

\begin{abstract}

    We present results from spectroscopic observations of three interacting galaxy systems, each associated with a primary galaxy — VCC 1748, SDSS J1106, and 2MASX J0834 — which have been proposed as potential examples of reverse morphological transformation from elliptical to spiral types driven by external gas accretion. Using Gemini/GMOS longslit spectroscopy, we obtain spectra along longslits connecting the regions of interest between the two galaxies in each pair. We model the spatially binned spectra with the Penalized Pixel-Fitting (pPXF) method to derive velocity fields and radial profiles of gas-phase metallicity, stellar metallicity, and stellar population age. Among the three systems, only the VCC 1748 system provides spectra with sufficiently high signal-to-noise ratios (S/N) to reveal coherent velocity and metallicity structures extending between the companion spiral galaxy and its elliptical target. The velocity profile along the gas-bridge of the system is smooth and continuous from the companion spiral arm to the target galaxy, and the elliptical target, VCC 1748, exhibits younger stellar populations and a high star formation rate (SFR) in its newly formed disk regions compared to its central parts. These findings indicate in-situ star formation at the newly-formed spiral arm of the elliptical target, consistent with ongoing gas inflow from the spiral companion into the elliptical target. Altogether, the results capture an intermediate stage of morphological evolution in which an originally quiescent elliptical galaxy begins to assemble a rotationally supported disk through interaction-driven gas accretion.

\end{abstract}


\section{Introduction} \label{sec:intro}

    Galaxy morphology is a key feature that encodes the diverse evolutionary pathways a system has experienced  \citep{Gunn_1972}. Morphology strongly correlates with global properties, such as luminosity and environmental conditions \citep{Dressler_1980, sandage_1985, park_2007}. These environmental influences include both external processes such as galaxy–galaxy interactions and ram-pressure stripping (e.g. \citealt{boselli2006,park2009,blanton2009}), and internal mechanisms including feedback from supernovae and active galactic nuclei (AGN; e.g. \citealt{werner2019}). Understanding galaxy morphology therefore requires detailed examination of both internal and external evolutionary factors. Over the past several decades, substantial evidence has accumulated establishing that major and minor mergers can drive the transformation of spiral galaxies into elliptical galaxies \citep{Barnes_1992, Burkert_2003, Hopkins_2008, Park_2008,hwang2009}.

    \begin{figure*}[ht!]
    \centering
    \includegraphics[width=1.0\textwidth]{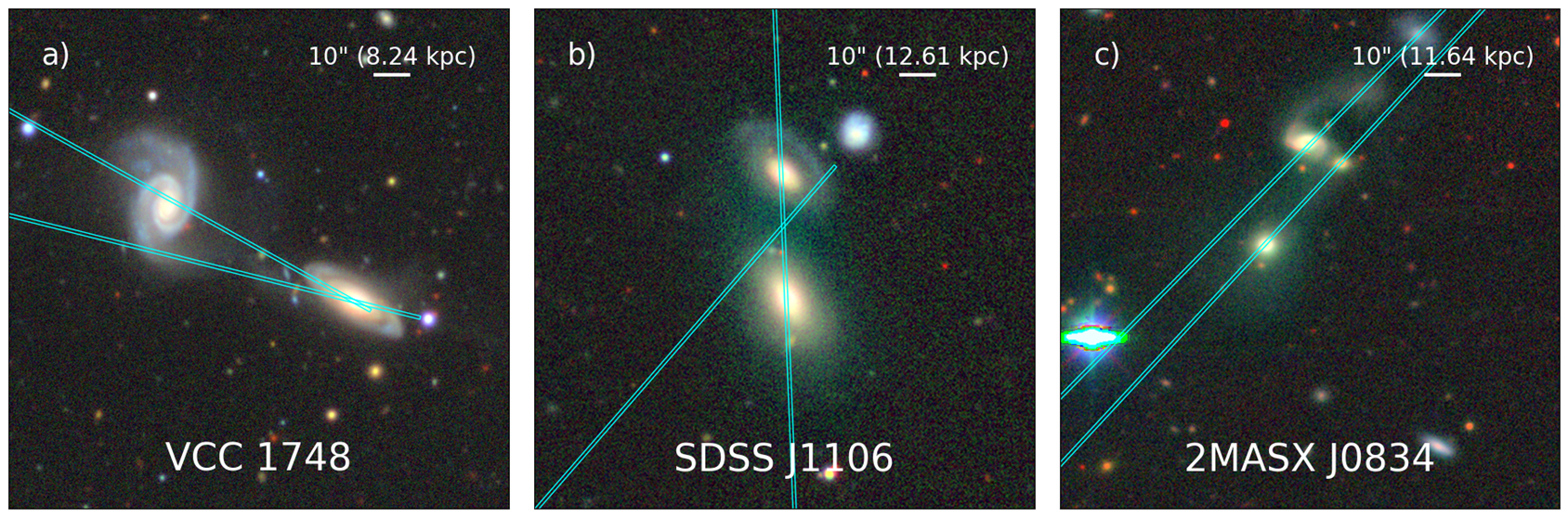}
    \caption{DESI images of the target systems with slit locations and pixel-sum centers overlaid. From left to right: (a) VCC 1748, (b) SDSS J1106, and (c) 2MASX J0834. Two slits are placed for each system: one intersecting the centers of both galaxies, and the other aligned with the observed gas–accretion feature.}
    \label{fig:photometric_img}
    \end{figure*}
    
    Conversely, several studies have suggested that, under specific conditions, elliptical galaxies may undergo a reverse morphological transformation, regenerating disk-like structures through external gas accretion. \cite{Park_2008} studied galaxies in the Sloan Digital Sky Survey Data Release 5 (SDSS DR5; \citealt{sdss_dr5}) and found that the galaxy morphology correlates with the morphology of the nearest neighbor when the pair separation is below the virial radius. In particular, systems such as VCC 1748, SDSS J1106, and 2MASX J0834 exhibit close pairs within the virial radius and display visible gas-transfer features, hinting at developing blue spiral structures forming around otherwise red elliptical galaxies (see Figure 4 of \citealt{Park_2008}). Based on the SDSS Mapping Nearby Galaxies at APO (MaNGA; \citealt{manga}), \cite{lee2023} reported that there are also hints of rejuvenated star-formation activity in elliptical galaxies, triggered by galaxy mergers or gas accretion from the cosmic web. These characteristics point to the possibility of reverse morphological transformation. 
    
    Related investigations further support this scenario. \cite{Hao_2019} analyzed red spiral galaxies in the SDSS MaNGA survey and argued that gas-rich major mergers can produce disk regrowth in otherwise quenched galaxies. \cite{Diaz_2018} performed hydrodynamical N-body simulations of mergers between elliptical and gas-rich spirals that evolve into S0 galaxies. Additionally, \cite{Park_2022} demonstrated using the TNG50 simulation \citep{TNG} that some quiescent galaxies can subsequently re-form disks. Collectively, these studies show that simulation-based galaxy evolution models allow red ellipticals to acquire disks under certain environmental conditions. 
    
    Motivated by these theoretical and observational hints, we conduct new spectroscopic observations with GMOS-N longslit to search for direct evidence of morphological transformation from elliptical to spiral galaxies driven by external gas accretion. To our knowledge, this work provides the first observational characterization of such an ongoing transformation. In Section \ref{sec:data}, we describe the observations and data-analysis methodology, including the derivation of key physical quantities from the spectra. Section \ref{sec:results} presents the velocity fields, age measurements, and stellar and gas metallicity profiles. Section \ref{sec:discussion} discusses the physical implications of these results for morphological evolution. Section \ref{sec:summary} summarizes our findings.

\section{Observations and Data} \label{sec:data}

\subsection{Observations}

    Our targets consist of three pairs of interacting galaxies selected from the study of \cite{Park_2008}, which are identified systems in SDSS DR5 exhibiting morphological disturbances and indications of gas transfer. Out of the six example systems presented in the work above, we select the three pairs with the smallest projected separations and the clearest visual evidence of a gas-accretion bridge feature. Observing DESI Legacy Imaging Surveys data \citep{DESI}, all three targets have projected separations within $\sim$20 $h^{-1}$ kpc. Additionally, each system shows a narrow low-surface-brightness structure connecting the two galaxies, suggesting cold-gas transfer between the galaxies. We label the targets in Figure \ref{fig:photometric_img} as (a) VCC~1748, (b) SDSS~J1106, and (c) 2MASX~J0834, following the names of the target elliptical galaxies.

    We conducted longslit spectroscopy using Gemini/GMOS-N (Program ID: GN-2023B-Q-112), with observations completed on December 15, 2023. A summary of the observational setup is provided in Table \ref{table:obs_summary}. The exposure time for each longslit configuration was $9\times450 \; \mathrm{sec}$. We employed the R831\_G5302 grating with the GG455\_G0305 order-blocking filter. The slit width was set to 1\arcsec{}, and the spatial axis was binned by two pixels. We applied both spectral and spatial dithering to mitigate chip gaps and improve flat-field quality.

    \begin{figure*}[ht!]
    \centering
    \includegraphics[width=0.95\textwidth]{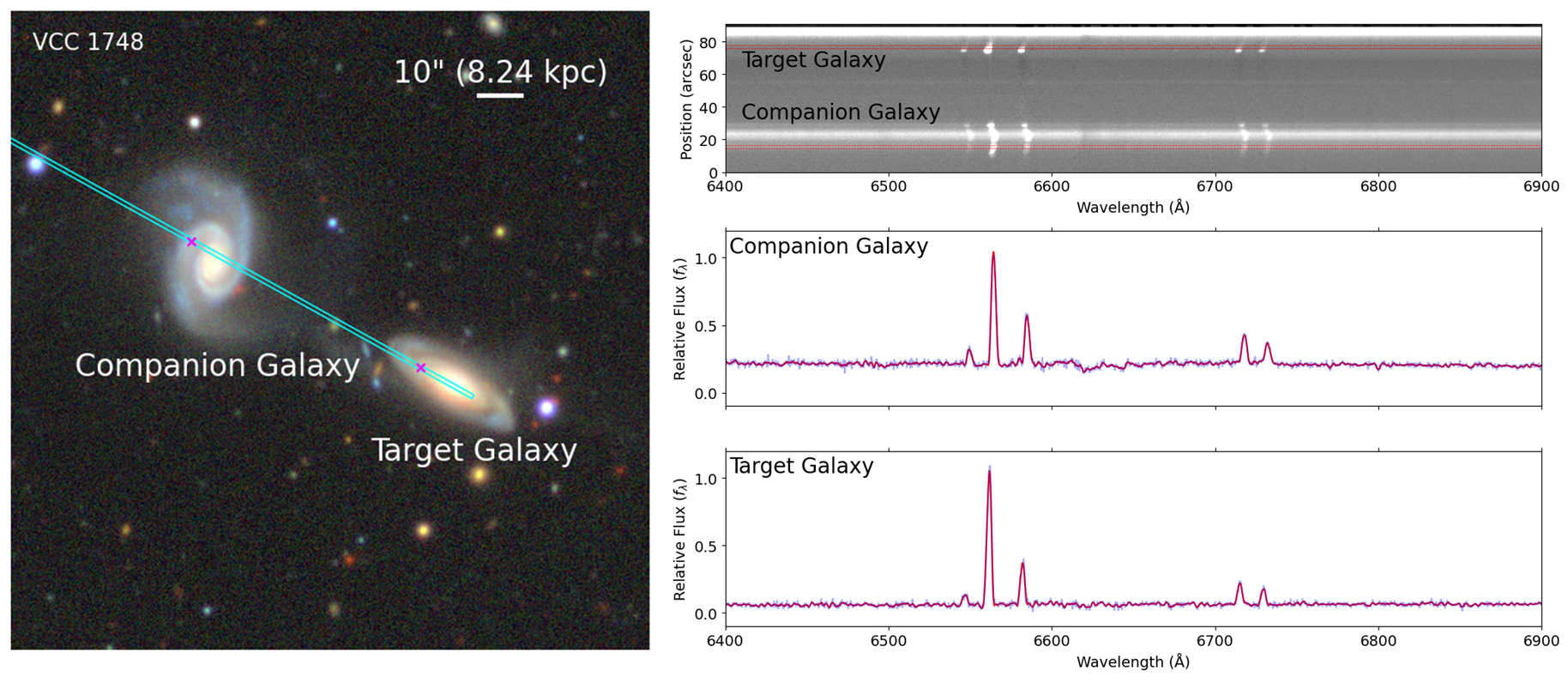}
    \caption{Images illustrating the data-reduction process. Left: DESI image of VCC~1748 showing the central slit and examples of pixel-summed regions. Upper right: flux- and wavelength-calibrated GMOS-N 2D spectrum. Middle and lower right: example pixel-summed spectra from the companion and target galaxy, respectively.}
    \label{fig:Reduction}
    \end{figure*}

    %% -- Added by ImSang -- %%
    \setlength{\arrayrulewidth}{0.6pt}
    \begin{deluxetable}{lllll}[!htbp]
        \tabletypesize{\scriptsize}
        \renewcommand{\arraystretch}{1.7}
        \renewcommand{\tabcolsep}{2.1mm}
        \tablecaption{Summary of Gemini/GMOS-N Longslit Observations}
        \label{table:obs_summary}
        \tablehead{\colhead{Target\tablenotemark{a}} & \colhead{RA} & \colhead{DEC} & \colhead{Obs. Date (UT)}}
        \startdata
        VCC 1748 & 12:38:13.9 & 07:48:57.6 & 2023 Dec 14/15 \\
        SDSS J1106 & 11:06:41.8 & 42:48:54.0 & 2023 Nov 17/18 \\
        2MASX J0834 & 08:34:06.2 & 46:26:20.4 & 2023 Oct 11/12 \\
        \hline
        \enddata
        \tablenotetext{a}{Target information corresponds to one galaxy in each interacting pair.}
    \end{deluxetable}
    %% -------------------- %%

    As illustrated in Figure \ref{fig:photometric_img}, two slit orientations were used for each system: (1) a ``central'' slit intersecting the nuclei of both galaxies, and (2) a ``gas-bridge'' slit aligned with the putative accretion feature. For 2MASX~J0834, the slits were placed parallel rather than intersecting both nuclei due to the presence of multiple unlabeled clumps in the optical images, which complicated the slit placement. In that case, each slit was positioned to intersect either the galaxy center or the gas-transfer feature to best capture the system geometry. The wavelength coverage spans 450–750 nm.

\subsection{Data Handling and Analysis Tools}

    We reduce all GMOS-N data using the DRAGONS pipeline \citep{dragons}, which performs bias subtraction, flat-fielding, wavelength, and flux calibration. We show the resulting flux and wavelength-calibrated 2D spectrum in the upper-right panel of Figure \ref{fig:Reduction}. 

    We perform sky subtraction using a 100-pixel summed region located adjacent to, but uncontaminated by, the galaxy emission along the slit. To analyze spatial variations, we divide each longslit into binned regions selected to balance spatial resolution and signal-to-noise ratio (S/N). High-S/N regions (typically within the galaxy disks) are summed over shorter pixel ranges, while low-S/N regions (e.g., in the gas bridge) require larger bins.

    Starting from the edge of the companion galaxy, we first apply fixed 10-pixel bins and measured the S/N of each extracted spectrum. Adjacent bins with S/N~$<1$ are iteratively merged until the combined spectrum exceeds this threshold. Regions that remain below S/N~$=1$ after merging are discarded as undetectable. For slits with only a small number of usable spectra after this process, we repeat the procedure, allowing overlapping bins to boost S/N and produce smoother spatial profiles.

    \begin{figure*}[ht!]
    \centering
    \includegraphics[width=0.95\textwidth]{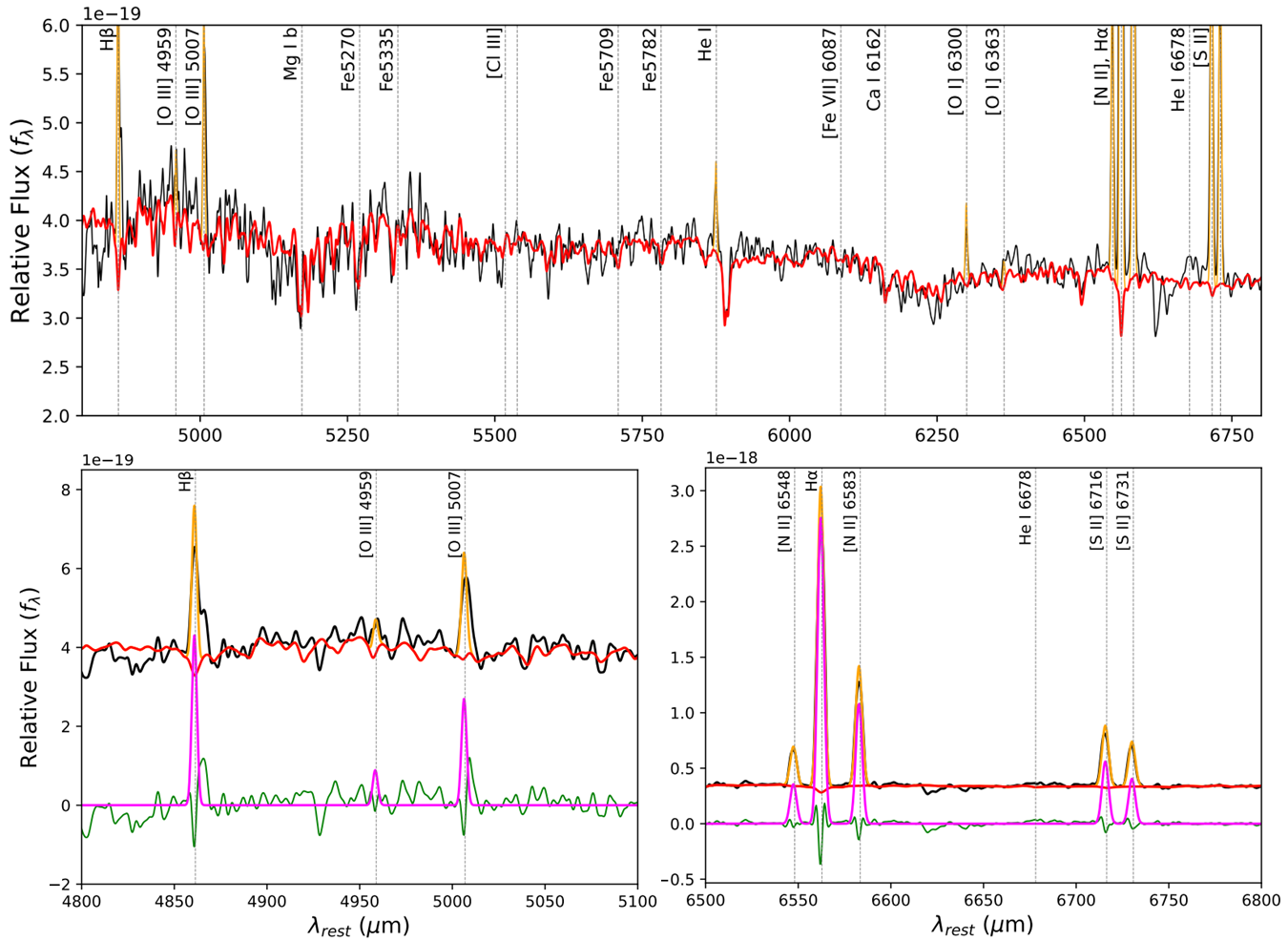}
    \caption{Examples of pPXF best-fit models for the VCC~1748 target galaxy. Black: observed spectrum; red: best-fit stellar continuum; orange: best-fit emission-line model; magenta: emission-only component; green: residuals. Top: full spectral range. Bottom: zoom-in of the [O\,III] region (left) and H$\alpha$+[S\,II] region (right).}
    \label{fig:pPXF}
    \end{figure*}

    We use the resulting spectra to derive physical parameters such as gas velocities, stellar/gas-phase metallicities, and luminosity-weighted ages with penalized PiXel-Fitting (pPXF) software \citep{ppxf} that allows fitting stellar continua and emission lines simultaneously. We derive gas velocities from the H$\alpha$ emission line, the strongest feature across all observable emission lines. We measure stellar metallicities and luminosity-weighted ages by fitting the stellar continuum with the E-MILES SSP templates \citep{emiles}. We compute gas-phase metallicities using the PP04 N2 diagnostic method \citep{PP04}, motivated by the limited availability of other strong emission lines such as [O\,\textsc{iii}] and H$\beta$ in several spectra. Equations (\ref{eq:N2}) and (\ref{eq:pp04}) describe the PP04 N2 diagnostic method to derive the gas-phase metallicity from the intensity ratio of the [N\,\textsc{ii}] and H$\alpha$ emission lines.

    \begin{equation}
    N2 \equiv \text{[N}_{\text{II}}]\ \lambda 6583 / \text{H}\alpha
    \label{eq:N2}
    \end{equation}
    \begin{equation}
    12 + \log(O/H) = 8.90 + 0.57 \times N2.
    \label{eq:pp04}    
    \end{equation}
    
    The pPXF returns independent stellar and gas velocity measurements, as well as emission line fluxes and continuum-weighted stellar population parameters as in Figure \ref{fig:pPXF}. Estimation of uncertainties on all pPXF-derived quantities is complete via 1000 Monte Carlo realizations of each spectrum. We retain only spectral bins with S/N~$>1$ to ensure robust fits.

\section{Results} \label{sec:results}

    As the goal of this project is to verify the ongoing morphological transformation in galaxy pairs driven by gas accretion from neighboring galaxies, we first examine whether material transfer occurs by analyzing the H$\alpha$ velocity profile along the gas-bridge slit. We then investigate whether the accreted gas is associated with in-situ transformation within the target elliptical galaxy by studying the spatial variations in gas metallicity, stellar metallicity, and luminosity-weighted stellar age.

    \begin{figure*}[h!]
    \centering
    \includegraphics[width=0.9\textwidth]{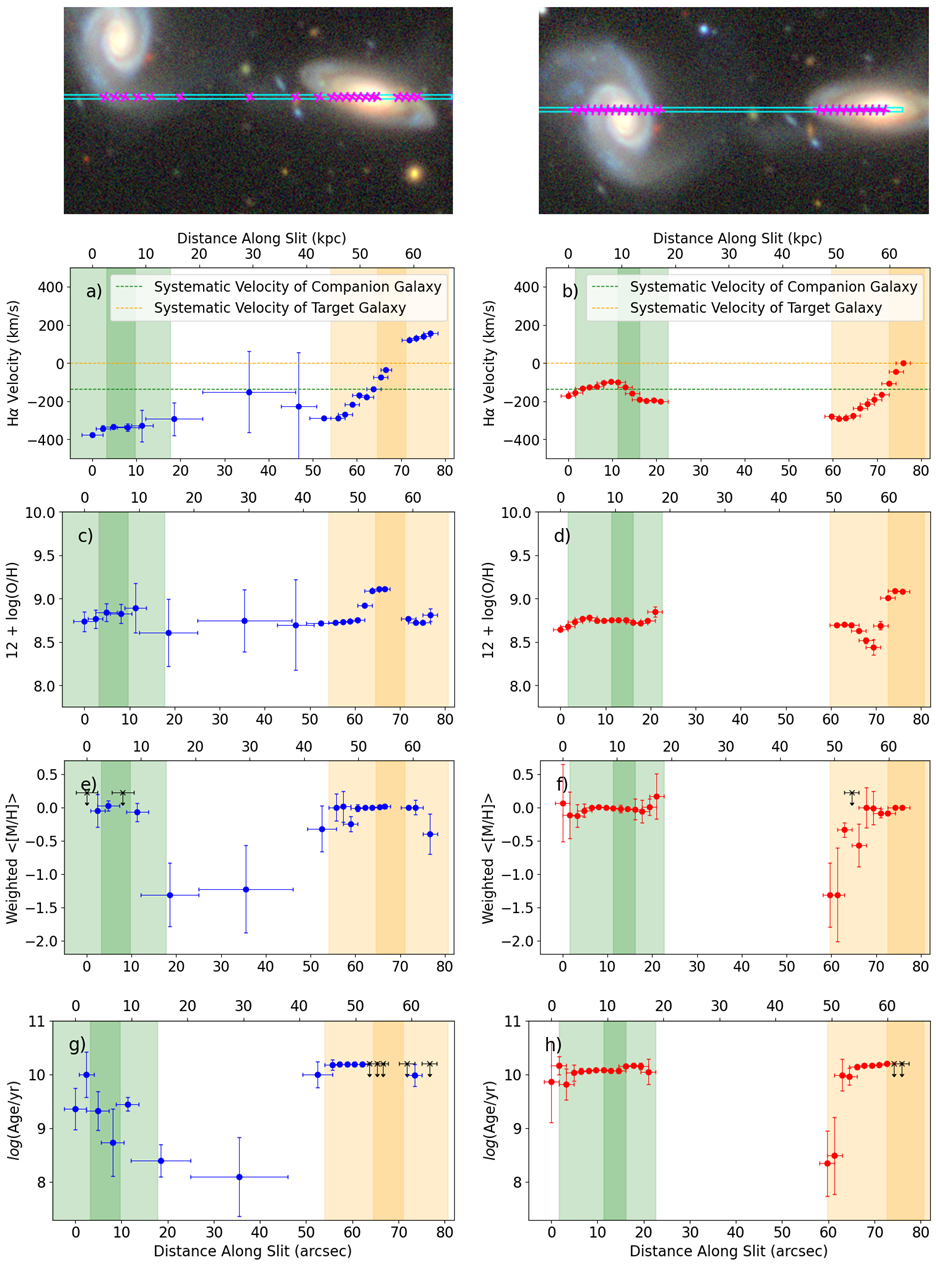}
    \caption{Pixel bins along the longslit of VCC 1748, with corresponding H$\alpha$ velocity, gas metallicity, stellar metallicity, and stellar age profiles from top to bottom. Green and yellow dashed line in H$\alpha$ velocity profile denotes velocity of companion and target galaxy center each, derived from the observed GMOS-N spectrum. Red points show the central slit; blue points show the gas-bridge slit. Green and yellow shaded regions mark the companion and target galaxy cores and disks. Black cross with downward arrows indicate upper limits imposed by template or fitting constraints.}
    \label{fig:VCC}
    \end{figure*}

    \begin{figure*}[ht!]
    \centering
    \includegraphics[width=0.9\textwidth]{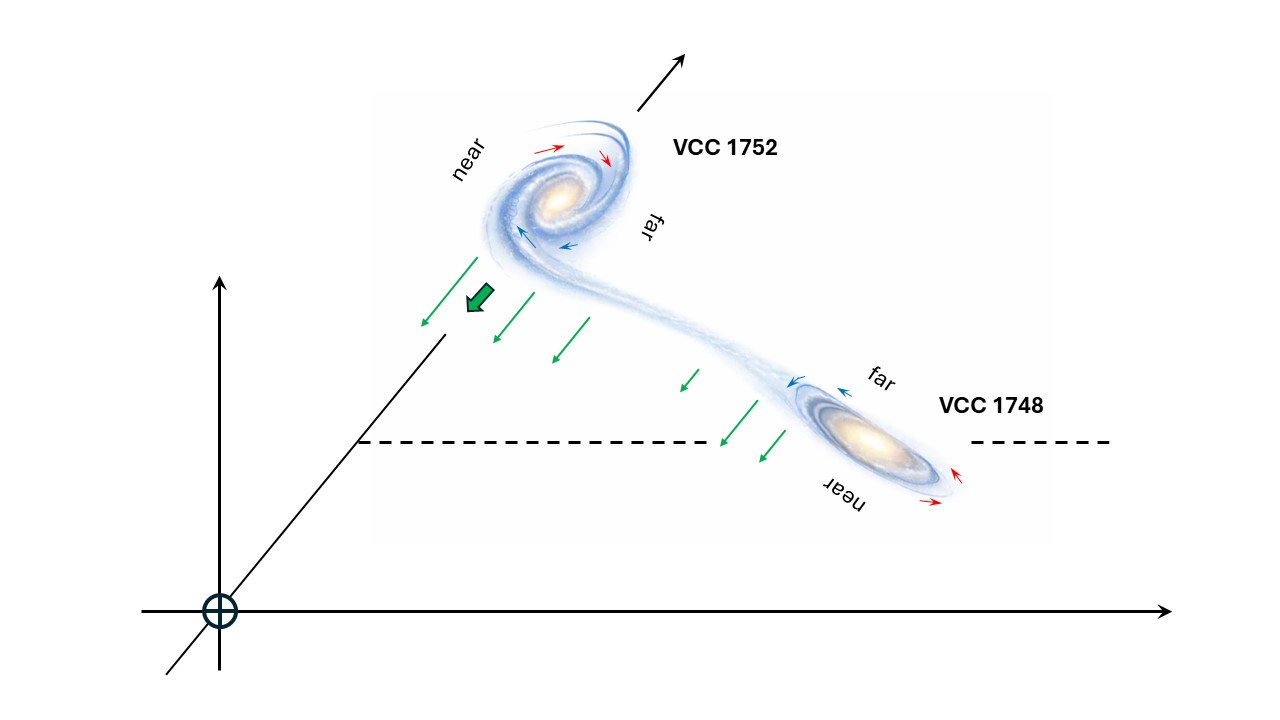}
    \caption{Schematic diagram of the interacting galaxy pair VCC 1748–VCC 1752. The diagram illustrates the relative locations and inclinations of the two galaxies. Their rotational directions are indicated by blue (approaching) and red (receding) arrows. Thin green arrows denote the line-of-sight gas velocities along the accretion (tidal) bridge. The thick arrow shows the relative radial velocity of VCC 1752 with respect to VCC 1748.}
    \label{fig:schematic}
    \end{figure*}

    We present the results for the most illustrative system, VCC~1748, which has spectra with sufficiently high S/N to study the region between paired galaxies (Figure \ref{fig:VCC}). The results for SDSS~J1106 and 2MASX~J0834 are provided in the Appendices \ref{app:sdss} and \ref{app:masx}. The red points in Figure \ref{fig:VCC} correspond to spectra extracted from the slit crossing both galaxy centers, while the blue points represent the gas-bridge slit. The green and yellow patches mark the approximate core and disk regions of the companion and target galaxy identified from color and surface-brightness transitions along the slit, respectively. Horizontal error bars represent the spatial width of each spectral bin. Figure \ref{fig:VCC} top row shows the center location of the data bin taken along the slit.

    The H$\alpha$ velocity profiles (Figure \ref{fig:VCC}, second row) show the kinematics of the gas. The green and yellow dashed lines mark the relative velocity of the companion and target galaxy measured from the H$\alpha$ velocity of each galaxy center, respectively. All velocities are plotted as the relative velocity to the target galaxy center H$\alpha$ velocity. In the central slit (second row, right panel b), both galaxies exhibit the expected rotational signatures across their disks. In contrast, the gas-bridge slit (second row, left panel a) reveals a continuous, monotonic velocity gradient connecting the spiral companion and the elliptical target. The companion's spiral arm shows velocities near $\sim-300$ km\,s$^{-1}$, rising to $\sim-200$ km\,s$^{-1}$ in the bridge and dipping again to $\sim-300$ km\,s$^{-1}$ near the target's disk before transitioning to the target’s rotation-dominated field.

    The gas metallicity profiles (Figure \ref{fig:VCC}, middle panels) show the measured gas metallicity using the H$\alpha$ and [NII] emission spectrum. In the central slit (middle row, right panel d), the nucleus of the elliptical target is more metal-rich (12+log(O/H)$\approx$9.08$\pm$0.05) than the nucleus of the spiral companion (8.75$\pm$0.04). The metallicity of the target disk (8.69$\pm$0.04) matches that of the outer disk of the companion, suggesting a shared, lower-metallicity gas reservoir. In the gas-bridge slit (middle row, left panel c), the companion’s disk (8.71$\pm$0.12) transitions smoothly into the target's disk (8.73$\pm$0.02), with a pronounced increase towards the target’s core (9.11$\pm$0.05). This continuity supports the physical connection inferred from the velocity field.

    The stellar metallicity profiles (Figure \ref{fig:VCC}, fourth row) reveal additional dynamical and evolutionary information. In the central slit (fourth row, right panel f), both galaxy cores exhibit similar metallicities ($\langle$[M/H]$\rangle \approx -0.02\pm0.06$), but stellar metallicity drops by $\sim$1.5 dex in the new spiral arm of the target. As with the gas metallicity, the stellar metallicity increases toward the side of the companion disk nearest to the elliptical target. As a possible explanation, we may suggest that the metal-rich stars within the cores get pulled out towards the pair galaxy side due to the gravitational interaction, causing a subtle increase in the metallicity near the paired side of the galaxy \citep{Kewley_2010}. In the gas-bridge slit (fourth row, left panel e), the stellar metallicity declines sharply (by $\sim$1.7 dex) along the bridge before rising again toward the target’s disk, suggesting the presence of young metal-poor stars transported from the companion’s outer spiral arm. The downward arrows mark the locations where the template-limited fits converge to the boundary value ($\sim$0.22), indicating the upper limit of the template rather than robust measurements.

    \begin{figure*}[ht!]
    \centering
    \includegraphics[width=0.9\textwidth]{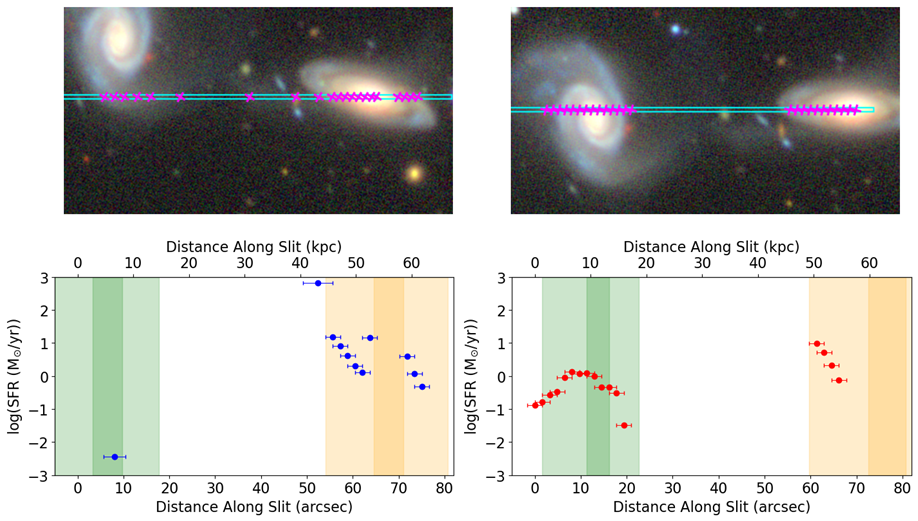}
    \caption{log(SFR) calculated with dust extinction for both slits of VCC 1748.}
    \label{fig:SFR}
    \end{figure*}
    
    Stellar population ages (Figure \ref{fig:VCC}, last row) show patterns similar to those of stellar metallicity. In the central slit (last row, right panel h), the galaxy cores contain old populations (log(age/yr)$\approx$10.07$\pm$0.05), whereas the new spiral arm of the target shows a dramatic age decrease (log(age/yr)$\approx$8.35$\pm$0.61). In the gas bridge slit (last row, left panel g), the age decreases from the companion core (9.97$\pm$0.26) to the gas bridge (8.00$\pm$0.90), then rises again toward the target (10.19$\pm$0.01). Similar to the phenomenon in stellar metallicity, the flat region near the target disk likely results from template limits, indicated by upper-limit arrows.

\section{Discussion} \label{sec:discussion}

\subsection{Transfer of Materials}

    Figure \ref{fig:VCC} (second row, left panel a) shows that the velocity profile along the gas-bridge slit is smooth and continuous from the companion spiral arm to the target galaxy. Based on our Bayesian change-point analysis, the gas velocity profile strongly supports a kinematically continuous bridge. We first emphasize that our interpretation of mass transfer does not rely solely on this kinematic continuity. The existence of the tidal bridge itself already indicates that tidal forces have stripped gas from the companion galaxy, implying mass transfer between the interacting galaxies. The significance of the observed kinematic continuity is that it demonstrates the stripped gas forms a dynamically coherent tidal stream connecting the companion and the target galaxy. In a prograde tidal encounter, gas removed from the rotating disk largely preserves its angular momentum, producing a smooth and continuous velocity field along the tidal bridge. The observed gradual radial-velocity variation without abrupt discontinuities is therefore consistent with gas originating from the companion galaxy and being transferred toward the target galaxy, as expected from numerical simulations of prograde tidal encounters \citep{Toomre_1972}. However, we also think that kinematic continuity by itself cannot uniquely distinguish ongoing gas transfer from a passive tidal bridge produced by the same interaction. We therefore consider additional observational evidence based on the system geometry, the stellar age distribution, the star formation rate, and the H$\alpha$ velocity dispersion.
    
    Figure \ref{fig:schematic} presents a schematic view of the interacting galaxy pair. Assuming that both galaxies have prograde spiral arms, the radial velocity profiles allow us to determine the near side of each galaxy. The companion spiral galaxy is left-handed, with its lower arm moving perpendicularly and the right-hand side approaching. The target galaxy is right-handed, with its left wing approaching. The blue and red arrows indicate the sense of rotation. The companion is approaching the target in the prograde orbit as the radial velocity of VCC1752 and the left wing of VCC1748 are both negative. The green arrows represent the line-of-sight gas velocities measured along the gas bridge, which trace the observed kinematics rather than the disk rotation field directly. The observed velocity pattern along the bridge may therefore differ from the naive rotational expectation of VCC 1752, as galaxy–galaxy interactions can significantly perturb the gas kinematics, and ambiguity in the inclination orientation of the disk (i.e. whether the near side of VCC 1752 faces the top or bottom of the image) could further alter the projected line-of-sight velocity pattern. \citep{Barton_1999, Repetto_2010}.
    
    \begin{figure*}[ht!]
    \centering
    \includegraphics[width=0.9\textwidth]{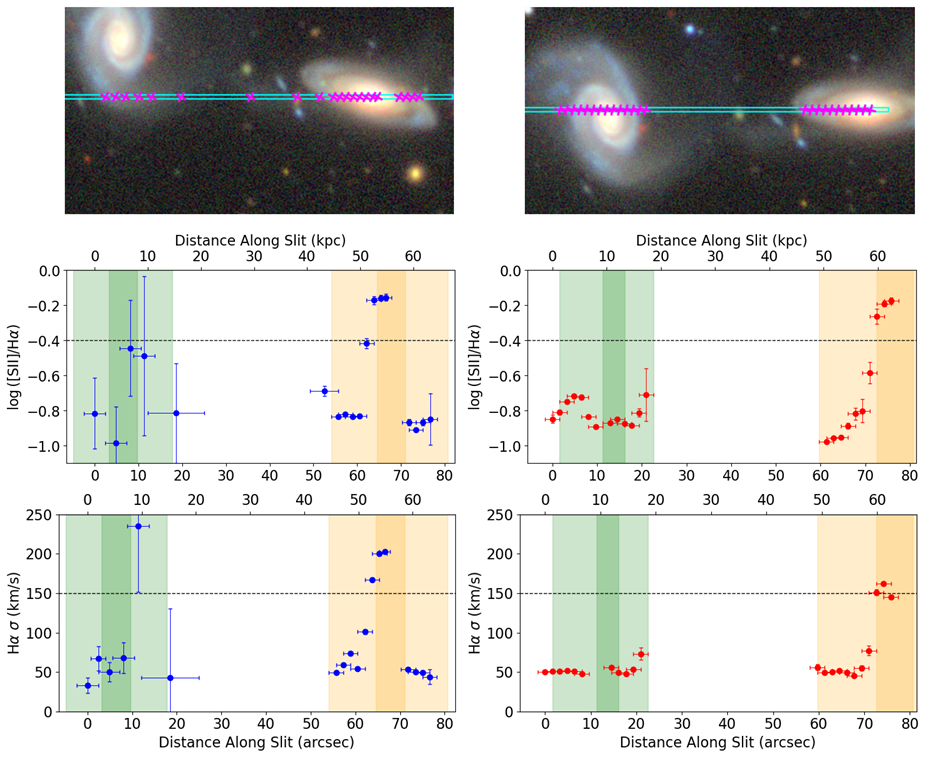}
    \caption{$\log$([S\,II]/H$\alpha$) and H$\alpha$ velocity dispersion for both slits of VCC~1748. The dashed line denotes the threshold separating shock-dominated and AGN narrow-line regions (NLR).}
    \label{fig:SII}
    \end{figure*}
    
    To further explain the schematics, the galaxy pair displays a systematic velocity difference of 150 km\,s$^{-1}$, measured from the central velocities observed in this program. Taking into account the measurement errors, this difference is statistically significant. If we assume that the two galaxies are currently interacting as indicated by the optical image, these can mean that the two galaxies are approaching each other in the line-of-sight direction; two galaxies are not purely perpendicular to the line-of-sight but rather placed diagonally with the target closer to us than the companion as depicted in Figure \ref{fig:schematic}. From this we may argue that line-of-sight gas velocity not only traces the shear/rotation of the connected system but also traces the actual gas movement along the bridge.
    
    Based on the SFR profiles, other research suggests that an accreting active gas bridge shows an enhancement in SFR at the point of interaction due to active gas accretion \citep{Moreno_2019, Sparre_2022}. Likewise, based on Figure \ref{fig:SFR}, in the VCC 1748 system, we may notice an enhancement in the SFR along the gas trail regions. This serves as a strong indicator of an ongoing gas accretion within the system. Consistent with this, the luminosity-weighted stellar ages along the companion spiral arm and the intervening gas trail are markedly young, generally $\log(\mathrm{Age/yr}) \lesssim 9$ and reaching $\log(\mathrm{Age/yr}) \sim 8$ near the bridge (Figure \ref{fig:VCC}, bottom row, left panel g), in contrast to the old stellar body of the target ($\log(\mathrm{Age/yr}) \sim 10$). Because these are luminosity-weighted ages, dominated by the most luminous and hence youngest populations, the presence of such young ages requires that luminous young stars be present at the epoch of observation; a population formed and then truncated in the distant past would instead have faded toward $\log(\mathrm{Age/yr}) \gtrsim 9.5$. The young ages therefore indicate recent, in-situ star formation along the trail rather than a purely passive tidal feature, in agreement with the SFR enhancement. Furthermore, Figure \ref{fig:SII} reveals a peak in the H$\alpha$ velocity dispersion near the 10–20 arcsec region. This enhancement likely reflects dynamical disturbance associated with the tidal interaction, possibly marking the disruption of the companion's outer spiral arm as material is stripped and enters the bridge. The overall dispersion structure remains consistent with an active gas-transfer stream connecting the two galaxies \citep{Zasov_2016}.
    
    The observed kinematic continuity indicates that the tidal bridge is a dynamically coherent structure connecting the companion and the target galaxy, rather than an unrelated gaseous feature. Although the relatively large spatial binning and associated fitting uncertainties in the gas-trail region ($\sim$20–50 arcsec) limit the statistical precision with which this continuity can be established, the observed velocity profile is unlikely to arise from random fluctuations alone. Moreover, the continuous velocity field is accompanied by consistent trends in gas metallicity, stellar metallicity, and stellar age across the bridge, all supporting a common physical origin for the stripped material. While kinematic continuity alone cannot distinguish between ongoing gas transfer and a recently formed passive tidal bridge, the combination of the bridge morphology, the line-of-sight geometry, the localized SFR enhancement, the young stellar populations, and the elevated H$\alpha$ velocity dispersion strongly favors the interpretation that the bridge represents an active gas-transfer stream produced during a prograde tidal interaction.
    We note that our data cannot formally exclude a fossil or tidal-relic scenario, in which part of the observed trail was deposited during an earlier stage of the encounter rather than being actively accreted at the present epoch. Such a scenario would not, however, alter the central conclusion of this work. The young ages along the trail place an upper bound of order a few $\times 10^{8}$ yr on the time since the associated star formation, so that any relic component must itself be recent and comparable to the orbital timescale of the encounter. More importantly, the inverse morphological transformation of the target is traced by the deposition of accreted gas and the resulting build-up of a young, rotationally supported, star-forming component, and this conclusion holds whether the transfer is strictly ongoing or partly relic in origin.
    
\subsection{Young Metal-poor Stars of Elliptical Target Disk}

    In the central slit shown in Figure \ref{fig:VCC} (right panels d, f, h), the region around 60–65 arcsec from the companion nucleus exhibits gas-phase metallicities comparable to those of the companion disk, while simultaneously showing markedly lower stellar metallicities and younger stellar ages. Beyond this region, the gas metallicity declines, whereas the stellar metallicity and stellar age increase toward the core of the target galaxy. Inspection of Legacy Survey images reveals that this low-metallicity, young stellar population spatially coincides with a disk-like structure in the target galaxy. Notably, a sharp increase in stellar metallicity and age occurs near a faint structural boundary separating the disk from the bulge component of the target.

    We consider three possible explanations for the $>$1.5 dex decrease in stellar metallicity observed at the target galaxy’s spiral arm near $\sim$60 arcsec in Figure~\ref{fig:VCC}f. First, the observed stellar population may be dominated by young accreted, metal-poor stars originating from the outskirts of the companion galaxy’s spiral arm and transported along the gas accretion trail. Second, metal-poor stars may have formed in situ within the spiral arm of the target galaxy. Third, ordinary in-situ star formation may have occurred in the spiral arm, but the stellar metallicity may be systematically underestimated due to observational or modeling limitations.

    The first scenario is unlikely, as although some delivery of metal-poor stars along the accretion trail is expected, it is improbable that such stars would be sufficiently numerous to dominate the measured stellar metallicity. Figure~\ref{fig:SFR} shows the SFR derived from dust-corrected H$\alpha$ emission, using the H$\beta$ line for extinction correction \citep{SFR}. Regions lacking reliable H$\beta$ measurements are excluded from the SFR calculation. The SFR in the newly formed spiral arm of the target galaxy peaks at nearly $10^{2.5}M_{\odot}\ \mathrm{yr}^{-1}$, approximately two orders of magnitude higher than the average SFR measured in both the target galaxy center and the companion galaxy. This indicates a significant episode of recent star formation in the spiral arm, arguing against a scenario in which accreted metal-poor stars dominate the stellar population.

    The second explanation, in-situ formation of metal-poor stars, is also disfavored. The gas-phase metallicity in this region is comparable to that of the target galaxy’s spiral arm and the central regions of the companion galaxy. If metal-poor stars had formed in situ, a corresponding decrease in gas-phase metallicity would be expected, which is not observed.

    The third explanation therefore provides the most plausible interpretation. Figure~\ref{fig:SFR} suggests that gas supplied via the accretion trail has fueled in-situ star formation at the outer edge of the target galaxy, producing a blue spiral arm embedded within an otherwise red elliptical system. We can explain the apparent depression in stellar metallicity by two related effects. First, in regions of intense star formation (Figure~\ref{fig:VCC}f, 60–65 arcsec), low continuum signal-to-noise ratios combined with strong nebular emission may compromise the measurement of stellar absorption features. Since pPXF derives stellar metallicities primarily from absorption lines, contamination by emission features can lead to systematically underestimated metallicities \citep{Fernandes_2005}. Second, the combination of low continuum S/N and age–metallicity degeneracy may drive the pPXF solution to the edge of the template grid, resulting in the lowest metallicity allowed by the stellar library. Indeed, the two lowest metallicity points in Figure~\ref{fig:VCC}f converge to an identical value of $\langle$[M/H]$\rangle = -1.31$. Such behavior is a characteristic of fits constrained by template limitations rather than physical extrema, suggesting that the apparent drop in metallicity is primarily an artifact of the fitting process \citep{Percival_2009, Ge_2018}.
    
\subsection{Nuclear Activity in an Elliptical Target Galaxy}

    \begin{figure*}[ht!]
    \centering
    \includegraphics[width=0.76\textwidth]{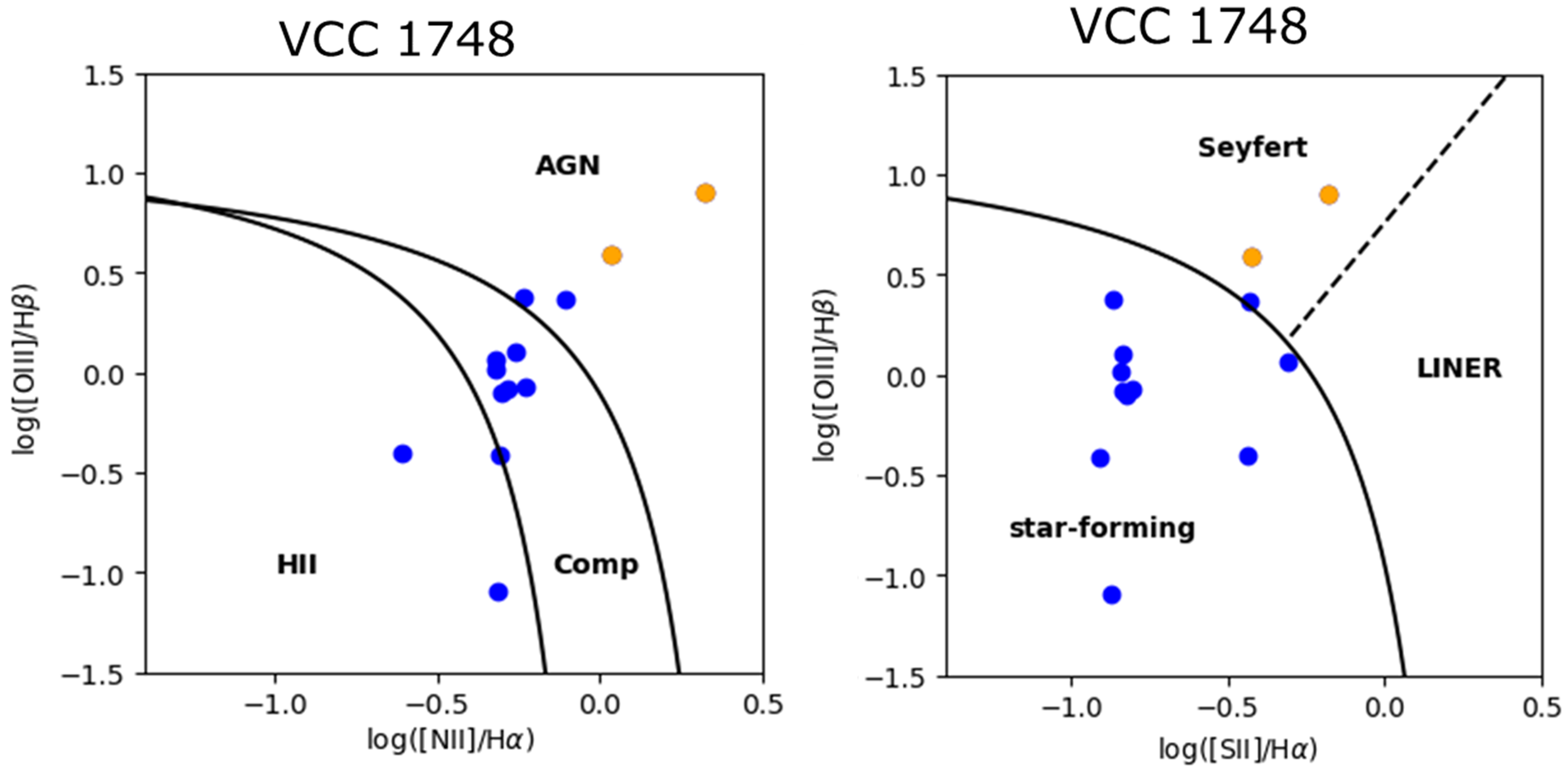}
    \caption{BPT diagrams for the gas-bridge slit of VCC~1748. Orange points represent the elliptical target galaxy's core; blue points mark all other regions. The target's nucleus lies in the AGN/Seyfert region in both diagrams.}
    \label{fig:BPT}
    \end{figure*}

    The sharp changes in gas metallicity observed near the target galaxy’s core can be explained by AGN-driven shock excitation. In Figure \ref{fig:VCC} (middle panels), the gas metallicity peaks sharply in the central region of the target galaxy by 0.5 dex. This enhancement is consistent with the shock excitation associated with AGN activity.

    AGN produce a hard ionizing spectrum, in contrast to the softer UV continuum of H\,II regions. Harder radiation increases the strength of forbidden lines such as [S\,II] and [O\,III] relative to Balmer lines, leading to elevated [S\,II]/H$\alpha$ ratios. We observe this behavior in Figures~\ref{fig:SII} and \ref{fig:BPT}. In Figure \ref{fig:SII} middle panels, the core of the target exhibits log([S\,II]/H$\alpha$) $> -0.4$, which exceeds the empirical threshold proposed by \cite{Ekaterina_2023} for shock-dominated excitation.

    In addition, the H$\alpha$ velocity dispersion in the target’s nucleus exceeds 150 km\,s$^{-1}$ (see also \citealt{Greene_2005, Mullaney_2013}), placing it firmly within the regime associated with AGN Narrow-Line Regions (NLR), as opposed to star-forming gas. Baldwin–Phillips–Terlevich (BPT) diagrams \citep{1981PASP...93....5B, 1987ApJS...63..295V} in Figure \ref{fig:BPT} further confirm that nuclear spectra fall in the AGN/Seyfert region, while the companion galaxy and gas-bridge points lie in the star-forming locus. Together, these diagnostics indicate that AGN-driven shocks in the target galaxy produce central metallicity enhancement and contribute to abrupt stellar-population transitions near the core.

\subsection{Asymmetry of the Spiral Arm Feature in SDSS J1106}

    \begin{figure}[!ht]
    \centering
    \includegraphics[width=0.37\textwidth]{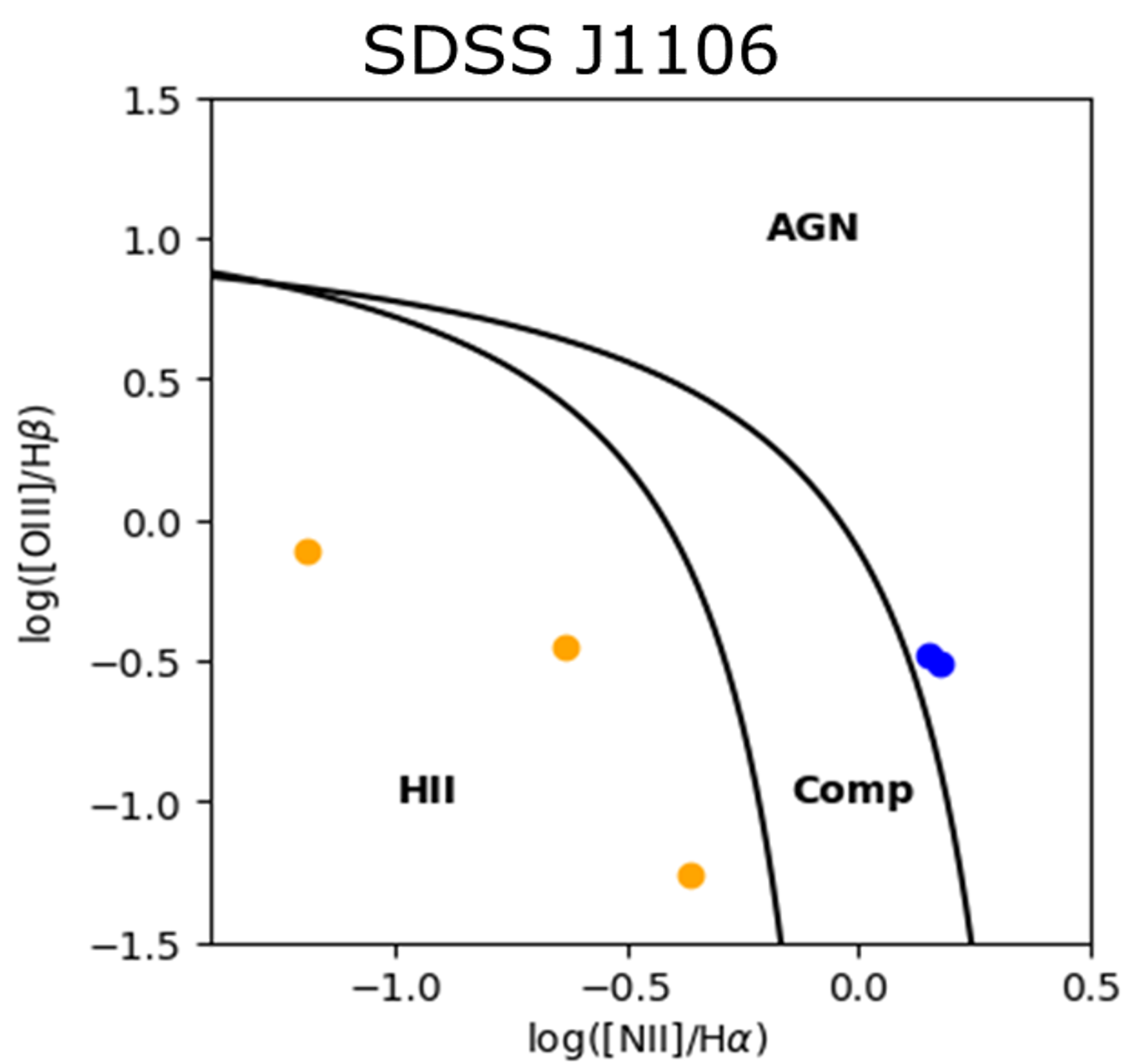}
    \caption{BPT diagrams comparing the spiral arm side and the opposite side without the spiral arm of SDSS J1106 companion galaxy. Orange points represent the region within the spiral arm; blue points mark the opposite region without the spiral arm.}
    \label{fig:SDSS_BPT}
    \end{figure}

    As discussed in Section~3.1 of \citet{Park_2008}, we will examine the asymmetric spiral arm observed in SDSS J1106, the middle panel of Figure~\ref{fig:photometric_img}. In the figure, a blue spiral arc is clearly visible on the upper side of the top companion galaxy, whereas no corresponding spiral structure is observed on the opposite side. \citet{Park_2008} proposed that the hot gas associated with the nearby elliptical galaxy may have ionized the spiral arm on the side facing the elliptical, ultimately leading to its dissipation. To investigate potential differences in the dominant ionization mechanisms between the two sides of the companion galaxy—one exhibiting a spiral arm and the other lacking such a feature—we construct a BPT diagram using emission-line measurements from the SDSS spectra presented in Appendix~\ref{app:sdss}.

    The resulting BPT diagram is shown in Figure~\ref{fig:SDSS_BPT}. The orange points denote spectra extracted from the side of the galaxy containing the spiral arm, while the blue points correspond to spectra from the opposite side without the spiral arm. We find that regions associated with the spiral arm predominantly occupy the H II-region locus, whereas regions on the opposite side fall mainly within the AGN-ionized regime. The BPT–[S II] diagnostic is not shown due to the insufficient number of reliable [S II] detections. Together, these diagnostics indicate that the dominant ionization source differs markedly between the two sides of the galaxy. This result suggests that the spiral arm on the right-hand side of the companion galaxy is sustained by ongoing star formation within H II regions, while the absence of a corresponding spiral arm on the opposite side may be linked to enhanced ionization driven by the neighboring elliptical galaxy, in qualitative agreement with the scenario proposed by \citet{Park_2008}.

\section{Summary} \label{sec:summary}

    We have investigated the ongoing morphological transformation of an elliptical galaxy into a spiral-like system induced by gas accretion from a neighboring target. Our three targets—VCC 1748, SDSS J1106, and 2MASX J0834—were selected from the list of the interacting galaxies in \cite{Park_2008} based on their clear signatures of the tidal interaction and visually identified gas-transfer features. Using Gemini/GMOS-N longslit spectroscopy, we derived gas velocity fields, gas and stellar metallicities, and luminosity-weighted stellar ages from pPXF fits to spatially binned spectra. Our main findings are summarized below.

    \begin{enumerate}
        \item[(1)] The H$\alpha$ velocity profile of VCC 1748 reveals a smooth and coherent flow of gas connecting the companion spiral to the target galaxy. Gas metallicity remains continuous along the bridge, while stellar age drops sharply by more than 1.5 dex, and SFR peaks by more than 3 dex on the outskirts of the target spiral arm. This pattern indicates that young in-situ star formation induced by accreted gas from the companion galaxy’s outer disk created an observable blue spiral arm to the elliptical target galaxy, inducing a morphology transformation from elliptical to spiral.
        
        \item[(2)] Measurements of log([S\,II]/H$\alpha$), the H$\alpha$ velocity dispersion and the BPT line ratios along the gas-bridge slit show clear evidence for an AGN NLR in the target galaxy. The associated shock excitation provides a natural explanation for the central enhancement in gas-phase metallicity, where AGN-driven shocks can elevate forbidden-line strengths and bias metallicity diagnostics toward higher values.

        \item[(3)] The asymmetry observed in the SDSS J1106 spiral arm of the top companion galaxy may originate from differences in the dominant ionization sources that affect the local star-forming gas. The side containing the spiral arm is likely ionized primarily by H II-region star formation, whereas the opposite side appears to be influenced by the left elliptical target. Such external ionization may suppress or evaporate pre-existing spiral structure, resulting in the morphological asymmetry observed in the companion galaxy of SDSS J1106.
    \end{enumerate}

    This study provides the first direct spectroscopic evidence of an ongoing morphological transformation in which an originally quiescent elliptical galaxy begins to rebuild a disk through external gas accretion. Although simulation-based studies and post-merger observations have long suggested the possibility of such reverse transformation, our results offer real-time observational confirmation. This work establishes an empirical foundation for exploring non-standard evolutionary pathways in galaxies and motivates future studies of interaction-driven morphological rejuvenation.

    % -------- Acknowledgments, Facilities, Software --------
    \begin{acknowledgments}
We thank the reviewer for the helpful comments that improved the original manuscript. HSH acknowledges support from the National Research Foundation of Korea (NRF) funded by the Korea government (MSIT; RS-2026-25482692) and the Global-LAMP Program funded by the Ministry of Education (RS-2023-00301976).
CBP is supported by the KIAS Individual Grant PG016904 at the Korea Institute for Advanced Study (KIAS) and by the National Research Foundation of Korea (NRF) grant funded by the Korean government (MSIT; RS-2024-00360385).
The work of HB was supported by the Basic Science Research Program through the National Research Foundation of Korea (NRF) funded by the Ministry of Education (RS-2025-25403440).
This work was supported by K-GMT Science Program (PID: GN-2023B-Q-112) of Korea Astronomy and Space Science Institute (KASI).
    \end{acknowledgments}

\facilities{Gemini(GMOS-N)}

\appendix

    Due to the low S/N of the data for SDSS J1106 and 2MASX J0834, we present these two systems separately below as it may be used as a reference for future works. We find statistically significant systematic line-of-sight velocity differences of $\approx$80 km\,s$^{-1}$ for SDSS J1106 and $\approx$200 km\,s$^{-1}$ for 2MASX J0834, indicating that the galaxies in each system are approaching along the line of sight, so that the observed line-of-sight gas velocities trace genuine intergalactic gas movement.

    \

\section{SDSS J1106} \label{app:sdss}

    We present the observational data for SDSS J1106 in Figure \sdss. In the H$\alpha$ velocity profiles, the central slit (second row, right panel b) traces a broadly continuous rotational profile across both galaxy cores, although large uncertainties in the disk regions rule out any robust kinematic interpretation. The bridge slit (second row, left panel a) lacks detected emission at several positions along the gas trail due to low S/N, making it difficult to identify a clear kinematic transition between the two galaxies. In the gas-phase metallicity profiles, the central slit (third row, right panel d) reveals metal-rich cores at a roughly uniform level (12+log(O/H)$\approx$9.1$\pm$0.23) with comparatively metal-poor disks ($\approx$8.7$\pm$0.31). The bridge slit (third row, left panel c) shows a companion-disk metallicity broadly consistent with both galaxy disks; as for VCC 1748, this consistency may hint at a physically connected gas structure despite the absence of detectable bridge emission. In the stellar metallicity profiles, the central slit (fourth row, right panel f) reveals metal-rich cores ($\langle$[M/H]$\rangle \approx 0.22\pm0.07$) and comparatively metal-poor disks ($\approx$-0.6$\pm$0.45~-1.6$\pm$0.94), though these values are dominated by upper limits from template or fitting constraints and must be treated with caution. The bridge slit (fourth row, left panel e) shows a roughly uniform stellar metallicity along the gas trail, consistent with both galaxy disks, and the stellar age profile follows an analogous trend.

    Several key differences from VCC 1748 emerge. The absence of detected bridge emission prevents us from tracing the velocity gradient or constraining the orbital geometry: whereas VCC 1748 shows a continuous, smooth H$\alpha$ velocity profile along the gas trail, SDSS J1106 shows a largely flat profile with no usable points in the trail region, making it difficult to confirm or refute gas accretion. The paucity of gas-metallicity points similarly prevents a definitive conclusion on connectivity, and the stellar metallicity and age profiles are dominated by upper limits along the trail, precluding detailed star-formation scenarios. Overall, SDSS J1106 lacks the spatial coverage and kinematic continuity needed to confirm ongoing accretion, and the present data provide only tentative evidence for a material connection along the gas bridge.

    \begin{figure*}[ht!]
    \centering
    \includegraphics[width=0.9\textwidth]{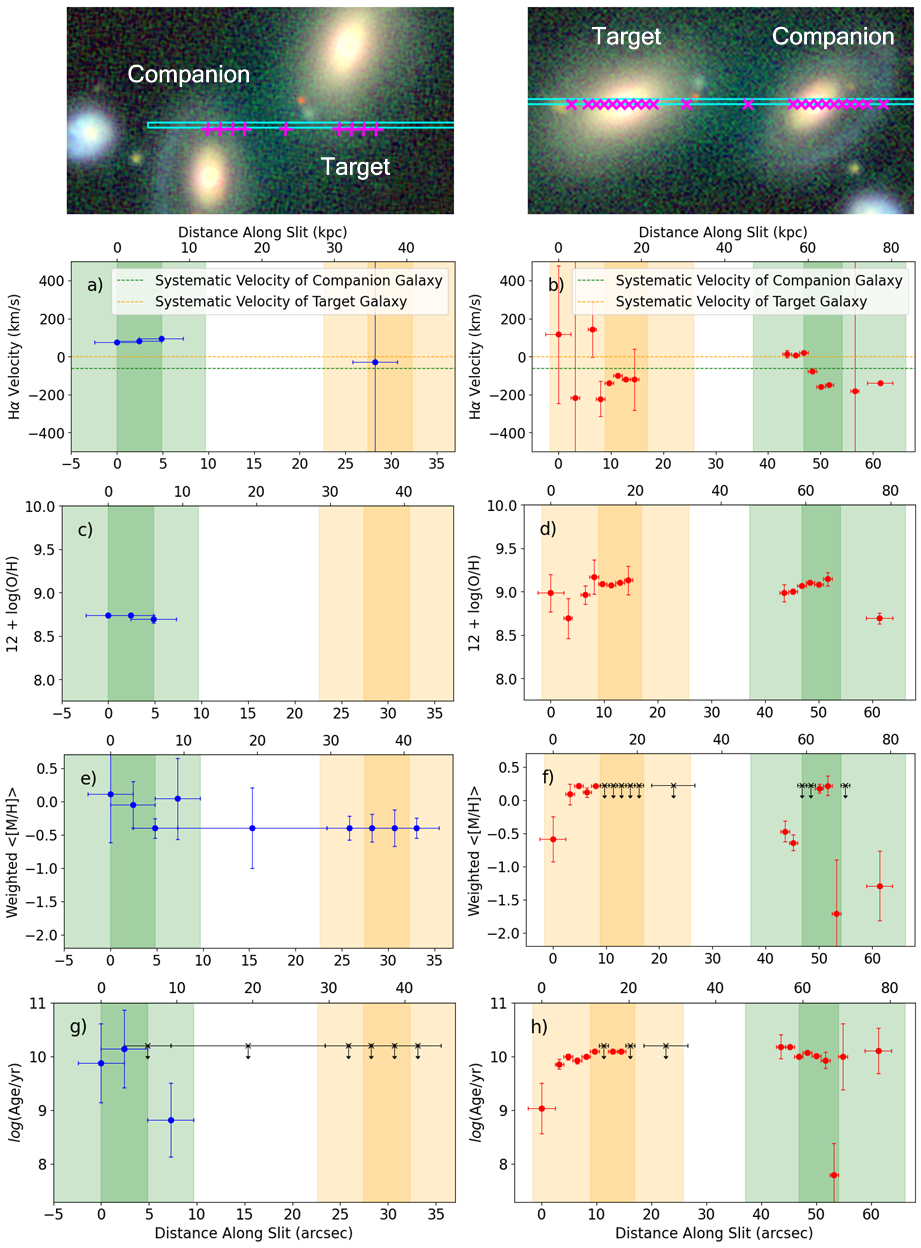}
    \caption{Image and plots of SDSS J1106 with the same format as Figure \ref{fig:VCC}.}
    \label{fig:SDSSJ}
    \end{figure*}

\section{2MASX J0834} \label{app:masx}

    We present the observational data for 2MASX J0834 in Figure \masx. Unlike the other systems, 2MASX J0834 was not observed with a central and bridge slit configuration; instead, two slits each pass through one galaxy nucleus and extend to the opposite galaxy's disk, so that both traverse the suspected intergalactic bridge. We refer to these as the blue and red slits. In the H$\alpha$ velocity profiles, the red slit (second row, right panel b) traces the rotational velocity of the target with emission extending toward the companion, while the blue slit (second row, left panel a) traces the companion with emission extending toward the target. Both slits recover the rotational kinematics of their respective galaxies, but several points carry large uncertainties, and the absence of detected emission at multiple bridge positions makes it difficult to assess ongoing accretion. In the gas-phase metallicity profiles, the red slit (third row, right panel d) yields too few reliable measurements to characterize the system, though the target disk metallicity ($\approx$8.65$\pm$0.24) is broadly consistent with the gas-trail metallicity ($\approx$8.68$\pm$0.92). The blue slit (third row, left panel c) shows a centrally elevated companion metallicity ($\approx$9.25$\pm$0.04) declining toward the disk ($\approx$8.81$\pm$0.08), with a gas-trail metallicity ($\approx$8.72$\pm$0.03) comparable to both disks, providing a tentative indication of a chemically continuous structure bridging the two galaxies. In the stellar metallicity profiles, the red slit (fourth row, right panel f) reveals metal-rich cores ($\langle$[M/H]$\rangle \approx -0.03\pm0.07$) and metal-poor disks ($\approx$-0.62$\pm$0.68), with anomalously low values at the outskirts of both galaxies that we attribute to the same systematic undershoot noted for VCC 1748. The blue slit (fourth row, left panel e) shows a roughly uniform stellar metallicity along the trail, consistent with both disks, and the stellar age profile follows an analogous trend.

    The key distinction from VCC 1748 is that the slit configuration covers the bridge only partially, so we cannot directly trace the bridge velocity gradient; combined with the large velocity uncertainties, this weakens the kinematic argument and the schematic analysis relative to VCC 1748. The gas metallicity likewise shows disconnected, unreliable points, although the chemical homogeneity along the bridge remains consistent with gas transported from the companion's outer disk toward the target, again analogous to VCC 1748 but on much weaker evidence. The stellar metallicity and age profiles show the same strongly depressed values along the bridge and at the disk edges as in VCC 1748, which may reflect either genuine metal-poor populations or systematic underestimation from modeling limitations. This degeneracy is testable: if future deeper observations reveal an elevated SFR in the metal-poor region, analogous to the star-forming spiral arm of VCC 1748, this would favor in-situ star formation driven by accreted gas, whereas a flat or declining SFR would instead indicate genuine metal-poor populations and little or no active accretion. As for SDSS J1106, the present data provide only tentative, limited evidence for a material connection along the gas bridge.

    Among the three systems, only VCC 1748 provides sufficient data quality to confirm ongoing gas accretion and in-situ star formation, supported by a continuous kinematic bridge, coherent metallicity gradients, and a localized SFR enhancement at the target's spiral arm. SDSS J1106 and 2MASX J0834 remain promising but unconfirmed candidates: their chemical homogeneity along the bridge is broadly consistent with the VCC 1748 scenario, yet the absence of detectable bridge emission and the dominance of upper limits in the stellar diagnostics prevent any definitive conclusion with the present data. A more definitive characterization of all three systems will require deeper, multi-wavelength observations.

    \begin{figure*}[ht!]
    \centering
    \includegraphics[width=0.9\textwidth]{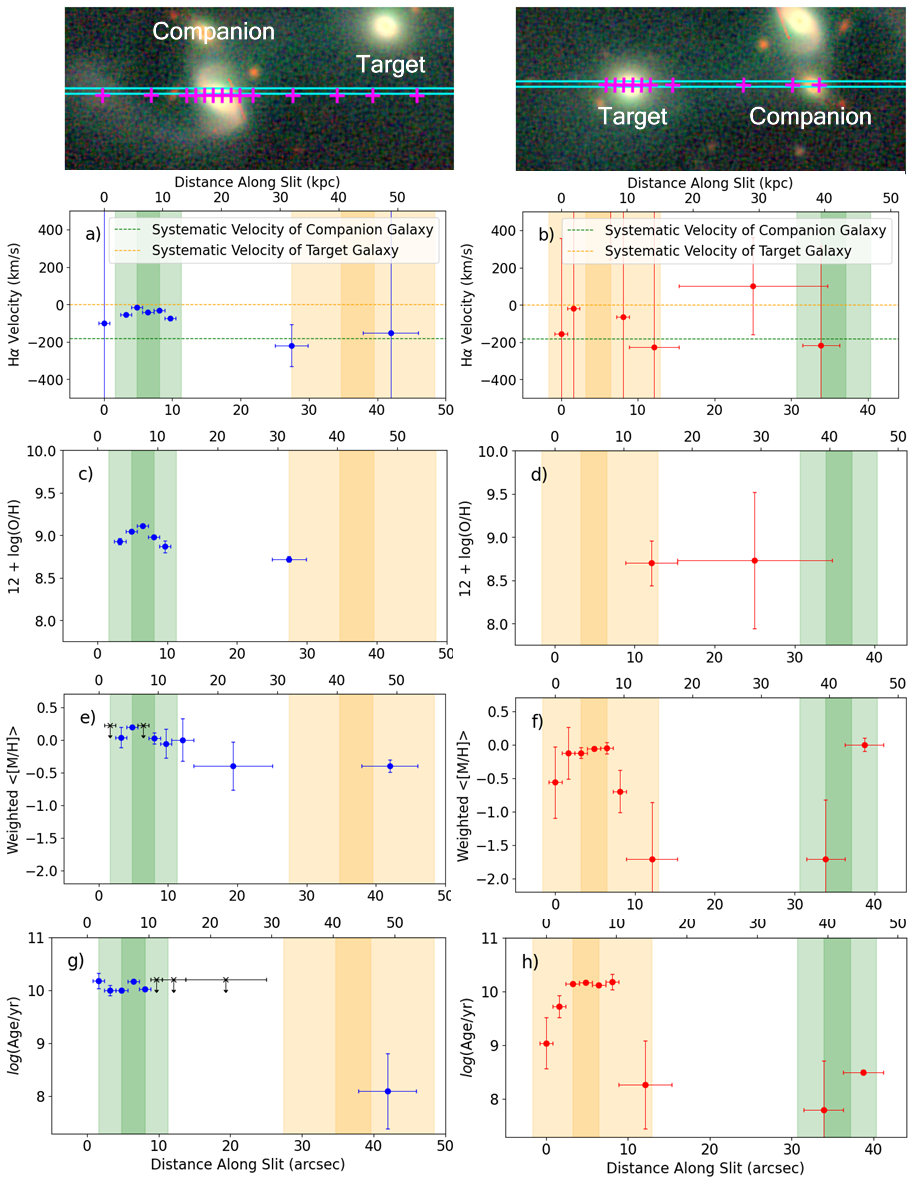}
    \caption{Image and plots of 2MASX J0834 with the same format as Figure \ref{fig:VCC}.}
    \label{fig:2MASXJ}
    \end{figure*}

\newpage
    
\bibliography{main}{}
\bibliographystyle{aasjournalv7}
\end{document}